\documentclass[a4paper, oneside, twocolumn, notitlepage, 10pt]{extarticle_ecoc}
\usepackage{ecoc}
\usepackage[helvet]{sfmath}
\usepackage[per-mode=symbol]{siunitx}
\DeclareSIUnit{\sample}{Sa}
\DeclareSIUnit{\baud}{Bd}
\DeclareSIUnit{\bit}{bit}
\DeclareSIUnit{\fourd}{4D}
\DeclareSIUnit{\eightd}{8D}
\DeclareSIUnit{\dBm}{dBm}
\DeclareSIUnit{\dB}{dB}
\DeclareSIUnit{\bps}{bps}

\usepackage{glossaries}
\glsdisablehyper
\newcommand{\SetCapsType}{normalcaps}
\usepackage{xspace}  
\usepackage[shortcuts]{extdash}
\usepackage{listofitems,pgffor}    
\usepackage{siunitx}
\usepackage{xstring}    
\usepackage{silence}
\usepackage{xparse}

\setsepchar{;}  

\ifdefined\silencecommonwarnings
\else
	\def\silencecommonwarnings{true} 
\fi

\ifbool{\silencecommonwarnings}{%
    \WarningFilter{ECOtools}{Cannot define: DH}%
    \WarningFilter{ECOtools}{Cannot define: PAM}%
    \WarningFilter{ECOtools}{Cannot define: QAM}%
    \WarningFilter{ECOtools}{Cannot define: SI}%
    \WarningFilter{ECOtools}{Cannot define: PV}%
    \WarningFilter{ECOtools}{Cannot define: LP}%
    \WarningFilter{ECOtools}{Cannot define: uLP}%
    \WarningFilter{ECOtools}{Redefining DH}%
    }{}

\makeatletter
\providecommand{\SetCapsType}{smallcaps}

\long\def\@scTrue{smallcaps}
\long\def\@scFalse{normalcaps}
\newcommand{\acroSCaps}[1]{%
    \ifx\SetCapsType\@scTrue 
        \textsc{#1}%
    \else
        \MakeUppercase{#1}%
    \fi
}
\makeatother

\usepackage{scalerel}
\makeatletter
\newcommand\scslash{%
\ifx\SetCapsType\@scTrue 
    \protect\stretchrel*{$/$}{\textsc{e}}
\else
    /
\fi
} 
\makeatother 

\makeatletter
\@ifpackageloaded{babel}{%
    \newcommand{\usuk}[2]{%
        \iflanguage{USenglish}{#1}{#2}%
    }%
}{%
    \newcommand{\usuk}[2]{%
        #1%
    }%
}%

\newcommand{\langcheck}[2]{
    \@ifpackageloaded{babel}{%
        \iflanguage{USenglish}{#1}{#2}%
    }{%
        #1%
    }%
}

\makeatother

\newcommand{\short}[1]{%
    \glsentrytext{#1}\xspace%
}
\newcommand{\Short}[1]{%
    \Glsentrytext{#1}\xspace%
}
\newcommand{\normal}[1]{%
    \gls{#1}\xspace%
}
\newcommand{\longacr}[1]{%
    \acrlong{#1}\xspace%
}
\newcommand{\plural}[1]{%
    \glspl{#1}\xspace%
}
\newcommand{\full}[1]{%
    \acrfull{#1}\xspace%
}
\newcommand{\fullplural}[1]{%
    \acrfullpl{#1}\xspace%
}
\newcommand{\Normal}[1]{%
    \Gls{#1}\xspace%
}
\newcommand{\Plural}[1]{%
    \Glspl{#1}\xspace%
}
\newcommand{\Full}[1]{%
    \Acrfull{#1}\xspace%
}
\newcommand{\Fullplural}[1]{%
    \Acrfullpl{#1}\xspace%
} 

\newcommand{\texpdfif}[2]{%
    \ifcsname texorpdfstring\endcsname%
        \texorpdfstring{#1{#2}}{#2\xspace}%
    \else%
        #1{#2}%
    \fi%
}

\newcommand{\checkanddefine}[3]{%
	\ifcsname #1\endcsname%
        \PackageWarning{ECOtools}{Cannot define: #1 already defined, trying to define g#1 instead.}%
        \ifcsname g#1\endcsname%
            \PackageWarning{ECOtools}{Cannot define: g#1 also already defined.}%
    	\else%
        	\expandafter\newcommand\csname g#1\endcsname{%
        	    \texpdfif{#2}{#3}%
    	    }%
        \fi%
	\else%
    	\expandafter\newcommand\csname #1\endcsname{%
    	    \texpdfif{#2}{#3}%
	    }%
    \fi%
}

\newcommand{\redefine}[3]{%
    \PackageWarning{ECOtools}{Redefining #1}%
	\expandafter\renewcommand\csname #1\endcsname{%
	    \texpdfif{#2}{#3}%
    }%
}

\newcommand{\nAcronym}[4][]{%
	\newacronym[#1]{#2}{#3}{#4}%
	\checkanddefine{s#2}{\short}{#2}%
	\checkanddefine{#2}{\normal}{#2}%
	\checkanddefine{l#2}{\longacr}{#2}%
	\checkanddefine{#2s}{\plural}{#2}%
	\checkanddefine{f#2}{\full}{#2}%
	\checkanddefine{f#2s}{\fullplural}{#2}%
	\checkanddefine{su#2}{\Short}{#2}%
	\checkanddefine{u#2}{\Normal}{#2}%
	\checkanddefine{u#2s}{\Plural}{#2}%
	\checkanddefine{fu#2}{\Full}{#2}%
	\checkanddefine{fu#2s}{\Fullplural}{#2}%
	\IfStrEq{#2}{DH}{
	    \redefine{#2}{\normal}{#2}%
	    }{}%
}%

\NewDocumentCommand\qam{g}{%
    \IfNoValueTF{#1}{%
        \texpdfif{\gls}{QAM}\xspace%
        }{%
        \StrLen{#1}[\stringlength]%
        \ifnum\stringlength=0%
            \texpdfif{\gls}{QAM}\xspace%
        \else%
            {\qamlisthelper{#1}}%
        \fi%
        }%
}

\let\QAM\qam

\DeclareRobustCommand\qamlisthelper[1]{%
    \readlist*\args{#1}%
    \acroSCaps{\args[1]\=/}%
    \ifnum\argslen = 2%
        { and \acroSCaps{\args[2]}\=/}%
    \fi%
    \ifnum\argslen > 2%
        \foreach \n in {2,...,\argslen}{%
            \ifnum\n = \argslen%
                {, and }%
            \else 
                {, }%
            \fi%
            {\acroSCaps{\args[\n]}\=/}%
        }%
    \fi%
    \ifglsused{QAM}%
        {}%
        {ary }%
    \texpdfif{\gls}{QAM}%
}%

\NewDocumentCommand\pam{g}{%
    \IfNoValueTF{#1}{%
        \texpdfif{\gls}{PAM}\xspace%
        }{%
        \StrLen{#1}[\stringlength]%
        \ifnum\stringlength=0%
            \texpdfif{\gls}{PAM}\xspace%
        \else%
            {\pamlisthelper{#1}}%
        \fi%
        }%
}

\DeclareRobustCommand\pamlisthelper[1]{%
    \readlist*\args{#1}%
    \ifglsused{PAM}{%
        \texpdfif{\gls}{PAM}%
        \acroSCaps{\=/\args[1]}%
        \ifnum\argslen = 2%
            { and \=/\acroSCaps{\args[2]}}%
        \fi%
        \ifnum\argslen > 2%
            \foreach \n in {2,...,\argslen}{%
                \ifnum\n = \argslen%
                    {, and }%
                \else%
                    {, }%
                \fi%
                {\=/\acroSCaps{\args[\n]}}%
            }%
        \fi%
    }{%
        \acroSCaps{\args[1]\=/}%
        \ifnum\argslen = 2%
            { and \acroSCaps{\args[2]}\=/}%
        \fi%
        \ifnum\argslen > 2%
            \foreach \n in {2,...,\argslen}{%
                \ifnum\n = \argslen%
                    {, and }%
                \else%
                    {, }%
                \fi
                {\acroSCaps{\args[\n]}\=/}%
            }%
        \fi%
        {ary }%
        \texpdfif{\gls}{PAM}%
    }%
}%

\NewDocumentCommand\lp{g}{%
    \IfNoValueTF{#1}{%
        \texpdfif{\normal}{LP}%
        }{%
        \StrLen{#1}[\stringlength]%
        \ifnum\stringlength=0%
            \texpdfif{\normal}{LP}%
        \else%
            \ifglsused{LP}{}{\texpdfif{\normal}{LP}\xspace}%
            \lplisthelper[lp]{#1}%
        \fi%
        }%
}
\let\LP\lp%

\NewDocumentCommand\ulp{g}{%
    \IfNoValueTF{#1}{%
        \texpdfif{\Normal}{LP}\xspace%
        }{%
        \StrLen{#1}[\stringlength]%
        \ifnum\stringlength=0%
            \texpdfif{\Normal}{LP}\xspace%
        \else%
            \ifglsused{LP}{%
                \lplisthelper[Lp]{#1}%
            }{%
                \texpdfif{\Normal}{LP}\xspace\lplisthelper[lp]{#1}%
            }%
        \fi%
        }%
}
\DeclareRobustCommand\lplisthelper[2][lp]{%
    \readlist*\args{#2}%
    \foreach \n in {1,...,\argslen}{%
        \ifnum \n > 1%
            \ifnum \argslen > 2%
                {, }%
            \else%
                { }%
            \fi%
        \fi%
        \ifnum \n = \argslen%
            \ifnum \argslen > 1%
                {and }%
            \fi%
        \fi%
        \ifnum \n = 1%
            {\acroSCaps{#1}}
        \else%
            {\acroSCaps{\MakeLowercase{#1}}}%
        \fi%
        {\textsubscript{\StrSplit{\args[\n]}{2}{\csA}{\csB}\acroSCaps{\csA}\csB}}
    }%
}%

\nAcronym{128SPQAM}{\acroSCaps{128-sp-16-qam}}{128-ary set-partitioning \QAM{16}}

\nAcronym{2A8PSK}{\acroSCaps{2a8psk}}{2-ary amplitude 8-ary phaseshift keying}

\nAcronym{3CCMCF}{\acroSCaps{3cc-mcf}}{3-core coupled-core multi-core fiber}

\nAcronym{4D}{\acroSCaps{4d}}{four-dimensional}
\nAcronym{4D64PRS}{\acroSCaps{4d-64prs}}{\usuk{four-dimensional 64-ary polarization-ring-switching}{four-dimensional 64-ary polarisation-ring-switching}}
\nAcronym{4DOS128}{\acroSCaps{4d-os128}}{four-dimensional orthant-symmetric 128-ary modulation format}

\nAcronym{5B4D2A8PSK}{\acroSCaps{5b4d-2a8psk}}{5-bit four-dimensional two-amplitude 8-ary phase-shift keying}

\nAcronym{6B4D2A8PSK}{\acroSCaps{6b4d-2a8psk}}{6-bit four-dimensional two-amplitude 8-ary phase-shift keying}

\nAcronym{7B4D2A8PSK}{\acroSCaps{7b4d-2a8psk}}{7-bit four-dimensional two-amplitude 8-ary phase-shift keying}

\nAcronym{8D}{\acroSCaps{8d}}{eight-dimensional}\nAcronym{8D2048PRS}{\acroSCaps{8d-2048prs}}{eight-dimensional 2048-ary polarization-ring-switching}
\nAcronym{8D2048PRST1}{\acroSCaps{8d-2048prs-t1}}{eight-dimensional 2048-ary polarization-ring-switching type 1}
\nAcronym{8D2048PRST2}{\acroSCaps{8d-2048prs-t2}}{eight-dimensional 2048-ary polarization-ring-switching type 2}
\nAcronym{8DAPSK}{\acroSCaps{8d-apsk}}{eight-dimensional amplitude-phase-shift keying}

\nAcronym{ABC}{\acroSCaps{abc}}{automatic bias controller}
\nAcronym{AC}{\acroSCaps{ac}}{alternating current}
\nAcronym{ADC}{\acroSCaps{adc}}{analog-to-digital converter}
\nAcronym{AIR}{\acroSCaps{air}}{achievable information rate}
\nAcronym{AO}{\acroSCaps{ao}}{adaptive optics}
\nAcronym{AOM}{\acroSCaps{aom}}{acoustic optical modulator}
\nAcronym{APD}{\acroSCaps{apd}}{avalanche photodiode}
\nAcronym{API}{\acroSCaps{api}}{application programming interface}
\nAcronym{AR}{\acroSCaps{ar}}{achievable rate}
\nAcronym{ARRWG}{\acroSCaps{a}rr\acroSCaps{wg}}{arrayed-waveguide grating}
\nAcronym{ASE}{\acroSCaps{ase}}{amplified spontaneous emission}
\nAcronym{ASK}{\acroSCaps{ask}}{amplitude-shift keying}
\nAcronym{ASIC}{\acroSCaps{asic}}{application-specific integrated circuit}
\nAcronym{ATS}{\acroSCaps{ats}}{alignment tracking sensor}
\nAcronym{AWG}{\acroSCaps{awg}}{arbitrary-waveform generator}
\nAcronym{AWGN}{\acroSCaps{awgn}}{additive white Gaussian noise}

\nAcronym{BBU}{\acroSCaps{bbu}}{baseband unit}
\nAcronym{BCH}{\acroSCaps{bch}}{Bose-Chaudhuri-Hocquenghem}
\nAcronym{BER}{\acroSCaps{ber}}{bit error rate}
\nAcronym{BERT}{\acroSCaps{bert}}{bit error rate tester}
\nAcronym{BICM}{\acroSCaps{bicm}}{bit-interleaved coded modulation}
\nAcronym{BMD}{\acroSCaps{bmd}}{bit-metric decoding}
\nAcronym{BPD}{\acroSCaps{bpd}}{balanced photo-diode}
\nAcronym{BPF}{\acroSCaps{bpf}}{bandpass filter}
\nAcronym{BPS}{\acroSCaps{bps}}{blind phase search}
\nAcronym{BPSK}{\acroSCaps{bpsk}}{binary phase-shift keying}
\nAcronym{BRGC}{\acroSCaps{brgc}}{binary reflected Gray code}
\nAcronym{BTB}{\acroSCaps{btb}}{back-to-back}

\nAcronym{CAGR}{\acroSCaps{cagr}}{compound annual growth rate}
\nAcronym{CCDM}{\acroSCaps{ccdm}}{constant composition distribution matching}
\langcheck{%
    \nAcronym{CCF}{\acroSCaps{ccf}}{coupled-core fiber}%
    }{%
    \nAcronym{CCF}{\acroSCaps{ccf}}{coupled-core fibre}%
}%
\nAcronym{CD}{\acroSCaps{cd}}{chromatic dispersion}
\nAcronym{CIR}{\acroSCaps{cir}}{channel impulse response}
\nAcronym{CMA}{\acroSCaps{cma}}{constant modulus algorithm}
\nAcronym{CMF}{\acroSCaps{cmf}}{core multiplicity factor}
\nAcronym{CMUX}{\acroSCaps{cmux}}{core multiplexer}
\nAcronym{COTS}{\acroSCaps{cots}}{commercial off-the-shelf}
\nAcronym{ChUT}{\acroSCaps{chut}}{channel under test}
\nAcronym[firstplural=channels under test (\acroSCaps{cut}s)]{CUT}{\acroSCaps{cut}}{channel under test}
\nAcronym{CPE}{\acroSCaps{cpe}}{carrier phase estimation}
\nAcronym{CPU}{\acroSCaps{cpu}}{central processing unit}
\nAcronym{CSPR}{\acroSCaps{cspr}}{carrier-to-signal power ratio}
\nAcronym{CUDA}{\acroSCaps{cuda}}{compute unified device architecture}
\nAcronym{CW}{\acroSCaps{cw}}{continuous wave}
\nAcronym{CCD}{\acroSCaps{ccd}}{charge-coupled device}

\nAcronym{DA}{\acroSCaps{da}}{driver amplifier}
\nAcronym{DAC}{\acroSCaps{dac}}{digital-to-analog converter}
\nAcronym{DC}{\acroSCaps{dc}}{direct current}
\nAcronym{DBP}{\acroSCaps{dbp}}{digital backpropagation}
\nAcronym{DCF}{\acroSCaps{dcf}}{\usuk{dispersion-compensating fiber}{dispersion-compensating fibre}}
\langcheck{%
    \nAcronym{DCI}{\acroSCaps{dci}}{data center interconnect}
    }{%
    \nAcronym{DCI}{\acroSCaps{dci}}{data centre interconnect}
}%
\nAcronym{DDLMS}{\acroSCaps{dd-lms}}{decision-directed least mean square}
\nAcronym{DEMUX}{\acroSCaps{demux}}{de-multiplexer}
\nAcronym{DFA}{\acroSCaps{dfa}}{\usuk{doped fiber amplifier}{doped fibre amplifier}}
\nAcronym{DFB}{\acroSCaps{dfb}}{distributed feedback}
\nAcronym{DGD}{\acroSCaps{dgd}}{differential group delay}
\nAcronym{DH}{\acroSCaps{dh}}{digital holography}
\nAcronym{DM}{\acroSCaps{dm}}{distribution matcher}
\nAcronym{DMR}{\acroSCaps{dm}}{dichroic mirror}
\nAcronym{DMA}{\acroSCaps{dma}}{direct memory access}
\nAcronym{DMD}{\acroSCaps{dmd}}{differential mode delay}
\nAcronym{DMG}{\acroSCaps{dmg}}{differential modal gain}
\nAcronym{DMGD}{\acroSCaps{dmgd}}{differential mode group delay}
\nAcronym{DML}{\acroSCaps{dml}}{directly-modulated laser}
\nAcronym{DP}{\acroSCaps{dp}}{\usuk{dual-polarization}{dual-polarisation}}
\nAcronym{DPC}{\acroSCaps{dpc}}{digital pre-compensation}
\nAcronym{DPE}{\acroSCaps{dpe}}{digital pre-emphasis}
\nAcronym{DPIQ}{\acroSCaps{dp-iqm}}{\usuk{dual-polarization \acroSCaps{iq}-modulator}{dual-polarisation \acroSCaps{iq}-modulator}}
\nAcronym{DPLL}{\acroSCaps{dpll}}{digital phase-locked loop}
\nAcronym{DQPSK}{\acroSCaps{dqpsk}}{differential quaternary phase-shift-keying}
\nAcronym{DRA}{\acroSCaps{dra}}{distributed Raman amplifier}
\nAcronym{DRE}{\acroSCaps{dre}}{digital resolution enhancer}
\nAcronym{DSB}{\acroSCaps{dsb}}{double-sideband}
\nAcronym{DSF}{\acroSCaps{dsf}}{\usuk{dispersion-shifted fiber}{dispersion-shifted fibre}} 
\nAcronym{DSO}{\acroSCaps{dso}}{digital sampling oscilloscope}
\nAcronym{DSP}{\acroSCaps{dsp}}{digital signal processing}
\nAcronym{DUT}{\acroSCaps{dut}}{device-under-test}
\nAcronym{DWDM}{\acroSCaps{dwdm}}{dense wavelength-division multiplexing}

\nAcronym{EAM}{\acroSCaps{eam}}{electro-absorption modulator}
\nAcronym{ECL}{\acroSCaps{ecl}}{external cavity laser}
\nAcronym{ED}{\acroSCaps{ed}}{Eucledian distance}
\nAcronym{EDF}{\acroSCaps{edf}}{\usuk{erbium-doped fiber}{erbium-doped fibre}}
\nAcronym{EDFA}{\acroSCaps{edfa}}{\usuk{erbium-doped fiber amplifier}{erbium-doped fibre amplifier}}
\nAcronym{ENOB}{\acroSCaps{enob}}{effective number of bits}
\nAcronym{ESS}{\acroSCaps{ess}}{enumerative sphere shaping}

\langcheck{%
    \nAcronym{FBG}{\acroSCaps{fbg}}{fiber Bragg grating}%
    }{%
    \nAcronym{FBG}{\acroSCaps{fbg}}{fibre Bragg grating}%
}%
\nAcronym{FD}{\acroSCaps{fd}}{frequency domain}
\nAcronym{FDE}{\acroSCaps{fde}}{\usuk{frequency domain equalizer}{frequency domain equaliser}}
\nAcronym{FEC}{\acroSCaps{fec}}{forward error correction}
\nAcronym{FFE}{\acroSCaps{ffe}}{\usuk{feed-forward equalizer}{feed-forward equaliser}}
\nAcronym{FFT}{\acroSCaps{fft}}{fast Fourier transform}
\nAcronym{FIR}{\acroSCaps{fir}}{finite impulse response}
\nAcronym{FLOPS}{\acroSCaps{flops}}{floating point operations per second}
\nAcronym{FMEDF}{\acroSCaps{fm-edf}}{\usuk{few-mode erbium-doped fiber}{few-mode erbium-doped fibre}}
\nAcronym{FMEDFA}{\acroSCaps{fm-edfa}}{\usuk{few-mode erbium-doped fiber amplifier}{few-mode erbium-doped fibre amplifier}}
\langcheck{%
    \nAcronym{FMF}{\acroSCaps{fmf}}{few-mode fiber}%
    }{%
    \nAcronym{FMF}{\acroSCaps{fmf}}{few-mode fibre}%
}%
\nAcronym[plural=FM-MCF, firstplural=\usuk{few-mode multi-core fibers}{few-mode multi-core fibres}]{FMMCF}{\acroSCaps{fm-mcf}}{\usuk{few-mode multi-core fiber}{few-mode multi-core fibre}}
\langcheck{%
    \nAcronym{FMPBGF}{\acroSCaps{fm-pbgf}}{few-mode photonic bandgap fiber}%
    }{%
    \nAcronym{FMPBGF}{\acroSCaps{fm-pbgf}}{few-mode photonic bandgap fibre}%
}%
\nAcronym{FOV}{\acroSCaps{fov}}{field of view}
\nAcronym{FPGA}{\acroSCaps{fpga}}{field-programmable gate array}
\nAcronym{FWM}{\acroSCaps{fwm}}{four-wave mixing}
\nAcronym{FSO}{\acroSCaps{fso}}{free-space optics}
\nAcronym{FUT}{\acroSCaps{fut}}{\usuk{fiber under test}{fibre under test}}

\nAcronym{GD}{\acroSCaps{gd}}{group delay}
\nAcronym{GI}{\acroSCaps{gi}}{graded-index}
\nAcronym{GFF}{\acroSCaps{gff}}{gain flattening filter}
\nAcronym{GIFMF}{\acroSCaps{gi-fmf}}{\usuk{graded-index few-mode fiber}{graded-index few-mode fibre}}
\nAcronym{GIMMF}{\acroSCaps{gi-mmf}}{\usuk{graded-index multi-mode fiber}{graded-index multi-mode fibre}}
\nAcronym{GMI}{\acroSCaps{gmi}}{\usuk{generalized mutual information}{generalised mutual information}}
\nAcronym{GNSE}{\acroSCaps{gnse}}{generalized nonlinear Schr\"{o}dinger equation}
\nAcronym{GPU}{\acroSCaps{gpu}}{graphics processing unit}
\nAcronym{GS}{\acroSCaps{gs}}{geometric shaping}
\nAcronym{GV}{\acroSCaps{gv}}{group velocity}
\nAcronym{GVD}{\acroSCaps{gvd}}{group velocity dispersion}
\nAcronym{GPIO}{\acroSCaps{gpio}}{general purpose input output}
\nAcronym{GUI}{\acroSCaps{gui}}{graphical user interface}

\langcheck{%
    \nAcronym{HCF}{\acroSCaps{hcf}}{hollow-core fiber}
    }{%
    \nAcronym{HCF}{\acroSCaps{hcf}}{hollow-core fibre}
}%
\nAcronym{HDFEC}{\acroSCaps{hd-fec}}{hard-decision forward error correction}
\nAcronym{HG}{\acroSCaps{hg}}{Hermite-Gaussian}
\nAcronym{HV}{\acroSCaps{hv}}{Hufnagel-Valley}
\nAcronym{HAP}{\acroSCaps{hap}}{Hufnagel-Andrew-Phillips}

\nAcronym{ICS}{\acroSCaps{ics}}{inter-core skew}
\nAcronym{ICXT}{\acroSCaps{ic-xt}}{inter-core cross-talk}
\nAcronym[plural=IL, firstplural=insertion losses (\acroSCaps{il})]{IL}{\acroSCaps{il}}{insertion loss}
\nAcronym{IFFT}{\acroSCaps{ifft}}{inverse fast Fourier transform}
\nAcronym{IMDD}{\acroSCaps{im}\scslash \acroSCaps{dd}}{intensity-modulation direct-detection}
\nAcronym{IQM}{\acroSCaps{iqm}}{in-phase and quadrature modulator}
\nAcronym{ISI}{\acroSCaps{isi}}{inter-symbol interference}

\nAcronym{JGN}{\acroSCaps{jgn}}{Japan Gigabit Network}

\nAcronym{KK}{\acroSCaps{kk}}{Kramers-Kronig}
\nAcronym{KIT}{\acroSCaps{kit}}{Karlsruhe Institute of Technology}

\nAcronym{LCOS}{\acroSCaps{LCoS}}{liquid crystal on silicon}
\nAcronym{LDPC}{\acroSCaps{ldpc}}{low-density parity-check}
\nAcronym{LEAF}{\acroSCaps{leaf}}{\usuk{large effective area fiber}{large effective area fibre}}
\nAcronym{LFSR}{\acroSCaps{lfsr}}{linear-feedback shift register}
\nAcronym{LG}{\acroSCaps{lg}}{Laguerre-Gaussian}
\nAcronym{LMS}{\acroSCaps{lms}}{least means squares}
\nAcronym{LLR}{\acroSCaps{llr}}{log-likelihood ratio}
\nAcronym{LO}{\acroSCaps{lo}}{local oscillator}
\nAcronym{LP}{\acroSCaps{lp}}{\usuk{linearly polarized}{linearly polarised}}
\nAcronym{LSPS}{\acroSCaps{lsps}}{\usuk{loop-synchronized polarization scrambler}{loop-synchronised polarisation scrambler}}
\nAcronym{LUT}{\acroSCaps{lut}}{lookup table}

\nAcronym{MB}{\acroSCaps{mb}}{Maxwell-Bolzmann}
\langcheck{%
    \nAcronym{MCF}{\acroSCaps{mcf}}{multi-core fiber}%
    }{%
    \nAcronym{MCF}{\acroSCaps{mcf}}{multi-core fibre}%
}%
\nAcronym{MDG}{\acroSCaps{mdg}}{mode dependent gain}
\nAcronym[firstplural=mode-dependent losses (\acroSCaps{mdl})]{MDL}{\acroSCaps{mdl}}{mode-dependent loss}
\nAcronym{MDM}{\acroSCaps{mdm}}{mode-division multiplexing}
\nAcronym{MEMS}{\acroSCaps{mems}}{micro-electro-mechanical systems}
\nAcronym{MF}{\acroSCaps{mf}}{matched filter}
\nAcronym{MFD}{\acroSCaps{mfd}}{mode field diameter}
\nAcronym{MI}{\acroSCaps{mi}}{mutual information}
\nAcronym{MIMO}{\acroSCaps{mimo}}{multiple-input multiple-output}
\nAcronym{ML}{\acroSCaps{ml}}{machine learning}
\nAcronym{MMA}{\acroSCaps{mma}}{multi-modulus algorithm}
\nAcronym{MMEDF}{\acroSCaps{mmedf}}{\usuk{multi-mode erbium-doped fiber}{multi-mode erbium-doped fibre}}
\nAcronym{MMEDFA}{\acroSCaps{mmedfa}}{\usuk{multi-mode erbium-doped fiber amplifier}{multi-mode erbium-doped fibre amplifier}}
\nAcronym{MMF}{\acroSCaps{mmf}}{\usuk{multi-mode fiber}{multi-mode fibre}}
\nAcronym{MMSE}{\acroSCaps{mmse}}{minimum mean squared error}
\nAcronym{MP}{\acroSCaps{mp}}{minimum phase}
\nAcronym{MPLC}{\acroSCaps{mplc}}{multi-plane light converter}
\nAcronym{MRC}{\acroSCaps{mrc}}{maximum ratio combining}
\nAcronym{MSE}{\acroSCaps{mse}}{mean squared error}
\nAcronym{MUX}{\acroSCaps{mux}}{multiplexer}
\nAcronym{MZM}{\acroSCaps{mzm}}{Mach-Zehnder modulator}
\nAcronym{MZI}{\acroSCaps{mzi}}{Mach-Zehnder interferometer}

\nAcronym{NA}{\acroSCaps{na}}{numerical aperture}
\langcheck{%
    \nAcronym{NANF}{\acroSCaps{nanf}}{nested antiresonant nodeless fiber}%
    }{%
    \nAcronym{NANF}{\acroSCaps{nanf}}{nested antiresonant nodeless fibre}%
}%
\nAcronym{NF}{\acroSCaps{nf}}{noise figure}
\nAcronym{NGMI}{\acroSCaps{ngmi}}{\usuk{normalized generalized mutual information}{normalised generalised mutual information}}
\nAcronym{NLSE}{\acroSCaps{nlse}}{nonlinear Schr\"{o}ding equation}
\nAcronym{NN}{\acroSCaps{nn}}{neural network}
\nAcronym{NIC}{\acroSCaps{nic}}{network interface card}
\nAcronym{NICT}{\acroSCaps{nict}}{National Institute of Information and Communications Technology}
\nAcronym{NIR}{\acroSCaps{nir}}{near-infrared}
\nAcronym{NRZ}{\acroSCaps{nrz}}{non-return-to-zero}
\nAcronym{NZDSF}{\acroSCaps{nz-dsf}}{\usuk{non-zero dispersion-shifted fiber}{non-zero dispersion-shifted fibre}} 

\nAcronym{OAM}{\acroSCaps{oam}}{orbital angular momentum}
\nAcronym{OBTB}{\acroSCaps{obtb}}{optical back-to-back}
\nAcronym{OCT}{\acroSCaps{oct}}{outer cladding thickness}
\nAcronym{ODE}{\acroSCaps{ode}}{ordinary differential equation}
\nAcronym{ODL}{\acroSCaps{odl}}{optical delay line}
\nAcronym{OEO}{\acroSCaps{oeo}}{optical-electrical-optical}
\nAcronym{OFC}{\acroSCaps{ofc}}{Optical Fiber Communications Conference}
\nAcronym{OFDR}{\acroSCaps{ofdr}}{optical frequency-domain reflectometer} 
\nAcronym{OFDM}{\acroSCaps{ofdm}}{orthogonal frequency division multiplexing}
\nAcronym{OH}{\acroSCaps{oh}}{overhead}
\nAcronym{OMFT}{\acroSCaps{omft}}{optical multi-format transmitter}
\nAcronym{OOK}{\acroSCaps{ook}}{on-off keying}
\nAcronym{OPLL}{\acroSCaps{opll}}{optical phase-locked loop}
\nAcronym{OSA}{\acroSCaps{osa}}{\usuk{optical spectrum analyzer}{optical spectrum analyser}}
\nAcronym{OSNR}{\acroSCaps{osnr}}{optical signal-to-noise ratio}
\nAcronym{OTDR}{\acroSCaps{otdr}}{optical time-domain reflectometer}
\nAcronym{OTF}{\acroSCaps{otf}}{optical tunable filter}
\langcheck{%
    \nAcronym{OVNA}{\acroSCaps{ovna}}{optical vector network analyzer}%
    }{%
    \nAcronym{OVNA}{\acroSCaps{ovna}}{optical vector network analyser}%
}%
\nAcronym{OTG}{\acroSCaps{otg}}{optical turbulence generator}

\nAcronym{PAM}{\acroSCaps{pam}}{pulse-amplitude modulation}
\nAcronym{PAS}{\acroSCaps{pas}}{probabilistic amplitude shaping}
\nAcronym{PAPR}{\acroSCaps{papr}}{peak-to-average power ratio}
\nAcronym{PBC}{\acroSCaps{pbc}}{\usuk{polarization beam combiner}{polarisation beam combiner}}
\langcheck{%
    \nAcronym{PBGF}{\acroSCaps{pbgf}}{photonic bandgap fiber}%
    }{%
    \nAcronym{PBGF}{\acroSCaps{pbgf}}{photonic bandgap fibre}%
}%
\nAcronym{PBS}{\acroSCaps{pbs}}{polarization beam splitter}
\nAcronym{PC}{\acroSCaps{pc}}{physical contact}
\nAcronym{PCVD}{\acroSCaps{pcvd}}{plasma chemical vapor depostion}
\nAcronym{PCG}{\acroSCaps{pcg64}}{64-bit permuted congruential generator}
\nAcronym{PD}{\acroSCaps{pd}}{photodiode}
\langcheck{%
    \nAcronym{PDL}{\acroSCaps{pdl}}{polarization-dependent loss}
    }{%
    \nAcronym{PDL}{\acroSCaps{pdl}}{polarisation-dependent loss}
}%
\nAcronym{PDM}{\acroSCaps{pdm}}{\usuk{polarization-division multiplexing}{polarisation-division multiplexing}}
\langcheck{%
    \nAcronym{PER}{\acroSCaps{per}}{polarization extinction ratio}%
    }{%
    \nAcronym{PER}{\acroSCaps{per}}{polarisation extinction ratio}%
}%
\nAcronym{PIC}{\acroSCaps{pic}}{photonic integrated circuit}
\nAcronym{PL}{\acroSCaps{pl}}{photonic lantern}
\nAcronym{PMBPSK}{\acroSCaps{pm-bpsk}}{polarization-multiplexed binary phase-shift keying}
\nAcronym{PMQPSK}{\acroSCaps{pm-qpsk}}{polarization-multiplexed quaternary phase-shift keying}
\nAcronym{PM8QAM}{\acroSCaps{pm-8qam}}{polarization-multiplexed 8-ary quadrature amplitude modulation}
\nAcronym{PMD}{\acroSCaps{pmd}}{\usuk{polarization mode dispersion}{polarisation mode dispersion}}
\langcheck{%
    \nAcronym{PMF}{\acroSCaps{pmf}}{polarization-maintaining fiber}%
    }{%
    \nAcronym{PMF}{\acroSCaps{pmf}}{polarization-maintaining fibre}%
}%
\nAcronym{PMP}{\acroSCaps{pmp}}{phase-matching point}
\nAcronym{PNOB}{\acroSCaps{pnob}}{physical number of bits}
\nAcronym{PON}{\acroSCaps{pon}}{passive-optical network}
\nAcronym{PRBS}{\acroSCaps{prbs}}{pseudorandom bit sequence}
\nAcronym{PROFA}{\acroSCaps{profa}}{\usuk{pitch reducing optical fiber array}{pitch reducing optical fibre array}}
\nAcronym{PPM}{\acroSCaps{ppm}}{pulse-position modulation}
\nAcronym{PS}{\acroSCaps{ps}}{probabilistic shaping}
\langcheck{%
    \nAcronym{PSCF}{\acroSCaps{pscf}}{pure silica core fiber}%
    }{%
    \nAcronym{PSCF}{\acroSCaps{pscf}}{pure silica core fibre}%
}%
\nAcronym{PSD}{\acroSCaps{psd}}{power spectral density}
\nAcronym{PSF}{\acroSCaps{psf}}{point spread function}
\nAcronym{PSK}{\acroSCaps{psk}}{phase-shift keying}
\nAcronym{PSP}{\acroSCaps{psp}}{\usuk{principal states of polarization}{principal states of polarisation}}
\nAcronym{PSW}{\acroSCaps{psw}}{\usuk{polarization switch}{polarisation switch}}
\nAcronym{PID}{\acroSCaps{pid}}{proportional–integral–derivative}
\nAcronym{PV}{\acroSCaps{pv}}{process value}

\nAcronym{QAM}{\acroSCaps{qam}}{quadrature amplitude modulation}
\nAcronym{QPSK}{\acroSCaps{qpsk}}{quaternary phase-shift keying}
\nAcronym{QSM}{\acroSCaps{qsm}}{quasi-single-mode}

\nAcronym{RAM}{\acroSCaps{ram}}{random-access memory}
\nAcronym{RC}{\acroSCaps{rc}}{raised cosine}
\nAcronym{RCMF}{\acroSCaps{rcmf}}{relative core multiplicity factor}
\nAcronym{RF}{\acroSCaps{rf}}{radio frequency}
\nAcronym{RI}{\acroSCaps{ri}}{refractive index}
\nAcronym{RLS}{\acroSCaps{rls}}{recursive least squares}
\nAcronym{RRC}{\acroSCaps{rrc}}{root-raised-cosine}
\nAcronym{ROADM}{\acroSCaps{roadm}}{reconfigurable optical add-drop multiplexer}
\nAcronym{ROI}{\acroSCaps{roi}}{region of interest} 
\nAcronym{RZDBPSK}{\acroSCaps{rz-dbpsk}}{return-to-zero differential binary phase-shift keying} 
\nAcronym{RZDQPSK}{\acroSCaps{rz-dqpsk}}{return-to-zero differential quaternary phase-shift keying} 

\nAcronym{S2}{\acroSCaps{S\textsuperscript{2}}}{spatially and spectrally resolved}
\nAcronym{SA}{\acroSCaps{sa}}{simulated annealing}
\nAcronym{SamPerSym}{\acroSCaps{sps}}{samples per symbol}
\nAcronym{SBS}{\acroSCaps{sbs}}{stimulated Brillouin scattering}
\nAcronym{SCM}{\acroSCaps{scm}}{subcarrier multiplexing}
\nAcronym{SDFEC}{\acroSCaps{sd-fec}}{soft-decision forward error correction}
\nAcronym{SDM}{\acroSCaps{sdm}}{space-division multiplexing}
\nAcronym{SE}{\acroSCaps{se}}{spectral efficiency}
\nAcronym{SER}{\acroSCaps{ser}}{symbol error rate}
\nAcronym{SI}{\acroSCaps{si}}{step index}
\nAcronym{SIFMF}{\acroSCaps{si-fmf}}{\usuk{step-index few-mode fiber}{step-index few-mode fibre}}
\nAcronym{SISMF}{\acroSCaps{si-smf}}{\usuk{step-index single-mode fiber}{step-index single-mode fibre}}
\nAcronym{SLM}{\acroSCaps{slm}}{spatial light modulator}
\nAcronym{SMF}{\acroSCaps{smf}}{\usuk{single-mode fiber}{single-mode fibre}}
\nAcronym{SMUX}{\acroSCaps{smux}}{spatial multiplexer}
\nAcronym{SNR}{\acroSCaps{snr}}{signal-to-noise ratio}
\nAcronym{SOA}{\acroSCaps{soa}}{semiconductor optical amplifier}
\nAcronym[firstplural=\usuk{states of polarization (\acroSCaps{sop})}{states of polarisation (\acroSCaps{sop})}]{SOP}{\acroSCaps{sop}}{\usuk{state of polarization}{state of polarisation}}
\nAcronym{SPM}{\acroSCaps{spm}}{self-phase modulation}
\nAcronym{SPS}{\acroSCaps{sps}}{samples per symbol}
\nAcronym{SRS}{\acroSCaps{srs}}{stimulated Raman scattering}
\nAcronym{SSB}{\acroSCaps{ssb}}{single-sideband}
\nAcronym{SSBI}{\acroSCaps{ssbi}}{signal-signal beat interference}
\nAcronym{SSFM}{\acroSCaps{ssfm}}{split-step Fourier method}
\nAcronym{SSMF}{\acroSCaps{ssmf}}{\usuk{standard single-mode fiber}{standard single-mode fibre}}
\nAcronym{STAXT}{\acroSCaps{staxt}}{short-term average cross-talk}
\nAcronym{STL}{\acroSCaps{stl}}{swept tunable laser}
\nAcronym{SVD}{\acroSCaps{svd}}{singular value decomposition}
\nAcronym{SW}{\acroSCaps{sw}}{sequence-wise}
\nAcronym{SWI}{\acroSCaps{swi}}{swept wavelength interferometry}
\nAcronym{SP}{\acroSCaps{sp}}{setpoint}

\nAcronym{TD}{\acroSCaps{td}}{time domain}
\nAcronym{TDE}{\acroSCaps{tde}}{\usuk{time domain equalizer}{time domain equaliser}}
\nAcronym{TDM}{\acroSCaps{tdm}}{time-domain multiplexing}
\nAcronym{TDMSDM}{\acroSCaps{tdm-sdm}}{time-domain multiplexed space-division multiplexing}
\nAcronym{TE}{\acroSCaps{te}}{transverse electric}
\nAcronym{TEC}{\acroSCaps{tec}}{thermally-expanded-core}
\nAcronym{TIA}{\acroSCaps{tia}}{trans-impedance amplifier}
\nAcronym{TM}{\acroSCaps{tm}}{transverse magnetic}
\nAcronym{TH4D}{\acroSCaps{th-4d}}{time domain hybrid four-dimensional}
\nAcronym{TH4D2A8PSK}{\acroSCaps{th-4d-2a8psk}}{time-domain hybrid four-dimensional two-amplitude eight-phase-shift keying}

\nAcronym{VCSEL}{\acroSCaps{vcsel}}{vertical-cavity surface emitting laser}
\nAcronym{VHDL}{\acroSCaps{vhdl}}{\acroSCaps{Vhsic} Hardware Description Language}
\nAcronym{VOA}{\acroSCaps{voa}}{variable optical attenuator}

\nAcronym{WDM}{\acroSCaps{wdm}}{wavelength-division multiplexing}
\nAcronym{WFS}{\acroSCaps{wfs}}{wavefront sensor}
\nAcronym{WGA}{\acroSCaps{wga}}{weakly guiding approximation}
\nAcronym{WGN}{\acroSCaps{wgn}}{white Gaussian noise}
\nAcronym[firstplural=wavelength selective switches (\acroSCaps{wss}s)]{WSS}{\acroSCaps{wss}}{wavelength selective switch}
\nAcronym{WC}{\acroSCaps{wc}}{weakly coupled}
\nAcronym{WCMCF}{\acroSCaps{wcmcf}}{\usuk{weakly-coupled multi-core fiber}{weakly-coupled multi-core fibre}}

\nAcronym{XPM}{\acroSCaps{xpm}}{cross-phase modulation}
\langcheck{%
    \nAcronym{XPOLM}{\acroSCaps{xp}ol\acroSCaps{m}}{cross-polarization modulation}
    }{%
    \nAcronym{XPOLM}{\acroSCaps{xp}ol\acroSCaps{m}}{cross-polarisation modulation}
}%
\nAcronym{XT}{\acroSCaps{xt}}{cross-talk}

\nAcronym{3DWG}{\acroSCaps{3dwg}}{3D-waveguide} 

\usepackage{subcaption}

\usepackage[capitalise]{cleveref}
\usepackage[export]{adjustbox}

\begin{document}
\selectlanguage{english}    


\title{Alignment of Free-Space Coupling of Few-Mode Fibre to Multi-Mode Fibre using Digital Holography}%


\author{Menno~van~den~Hout\textsuperscript{(1,*)},
        Sjoerd~van~der~Heide\textsuperscript{(1)},
        Thomas Bradley\textsuperscript{(1)},
        Amado~M.~Velazquez-Benitez\textsuperscript{(1),(2)}\\
        Nicolas K. Fontaine\textsuperscript{(3)},
        Roland Ryf\textsuperscript{(3)},
        Haoshuo Chen\textsuperscript{(3)},
        Mikael Mazur\textsuperscript{(3)},
        Jose~Enrique~Antonio-L\'opez\textsuperscript{(4)},\\
        Juan~Carlos Alvarado-Zacarias\textsuperscript{(4)}, 
        Rodrigo~Amezcua-Correa\textsuperscript{(4)},
         Marianne~Bogot-Astruc\textsuperscript{(5)},\\
        Adrian~Amezcua~Correa\textsuperscript{(5)}, 
        Pierre~Sillard\textsuperscript{(5)}, 
        and Chigo~Okonkwo\textsuperscript{(1)}
}

\maketitle                  



    \begin{strip}
        \begin{author_descr}

            \textsuperscript{(1)} High Capacity Optical Transmission Laboratory, Electro-Optical Communications Group,\\
            Eindhoven University of Technology, the Netherlands
            \textsuperscript{(*)}{\,\uline{m.v.d.hout@tue.nl}}\\
            \textsuperscript{(2)} Instituto de Ciencias Aplicadas y Tecnolog\'{i}a, Universidad Nacional Auton\'{o}ma de M\'{e}xico, Circuito Exterior S/N, Ciudad Universitaria, 04510, Mexico City.\\
            \textsuperscript{(3)} Nokia Bell Labs, 600 Mountain Ave, New Providence, NJ 07974, USA\\
             \textsuperscript{(4)} CREOL, The College of Optics and Photonics, University of Central Florida, Orlando, 32816, USA
             \textsuperscript{(5)} Prysmian Group, 644 Boulevard Est, Billy Berclau, 62092 Haisnes Cedex, France
        \end{author_descr}
    \end{strip}

\setstretch{1.1}
\renewcommand\footnotemark{}
\renewcommand\footnoterule{}


\begin{strip}
  \begin{ecoc_abstract}
    Off-axis digital holography is used to align a few-mode fiber to a multi-mode fiber in a free-space optical setup. Alignment based on power coupling measurements alone cannot guarantee low mode-dependent loss. The proposed alignment method enables reliable fiber coupling with low mode-dependent loss and crosstalk. \textcopyright\,\,2022~The~Author(s)
  \end{ecoc_abstract}
\end{strip}


\section{Introduction}
The potential of \SDM to greatly increase optical fiber transmission data rates has been shown \cite{Puttnam2021}. In \MDM, a subset of \SDM, the spatial diversity of \FMFs and \MMFs is exploited by modulating data on fiber modes as independent spatial paths, increasing throughput with respect to conventional single-mode systems. Light in these fibers has a complex spatial distribution. Therefore, when coupling between \SDM components optimized using total coupled power as an optimization metric, certain fiber modes may be disproportionately affected, leading to increased impairments such as \MDL and \XT. Hence, the spatial properties of coupling should be taken into account, requiring characterization tools able to provide such insight.

A full description of the spatial distribution of light can be obtained using digital holographic measurements. Off-axis \DH measures the amplitude and phase for both polarizations of a free-space optical signal by recording the interference between the signal field and a flat-phase reference \cite{thesisSjoerdDH,demoOFC,Fontaine2019a,Mazur2019,Alvarado-Zacarias2020,digholo}. Subsequent analysis of the measured interference patterns can reveal important metrics for \SDM transmission systems such as \MDL and \XT.

In this work, we demonstrate the use of off-axis \DH for the alignment of free-space coupling of light between a \FMF and a \MMF. Coupling is evaluated at various fiber positions. At each position, the total coupled optical power is measured using a free-space power meter and a transfer matrix of the \SDM subsystem is measured using \DH, which is used to calculate \MDL and \XT. It is shown that only maximizing total coupled optical power does not provide adequate coupling and severe \MDL penalties of up to \SI{20}{dB} are observed. Therefore, to ensure reliable results, the spatial distribution of the light must be taken into account when coupling is optimized in \SDM systems. Off-axis \DH is demonstrated to provide the necessary measurements for reliable automated alignment of \SDM devices and subsystems.

\section{Experimental setup}
\begin{figure*}
    \centering
    \includegraphics[width=0.99\linewidth]{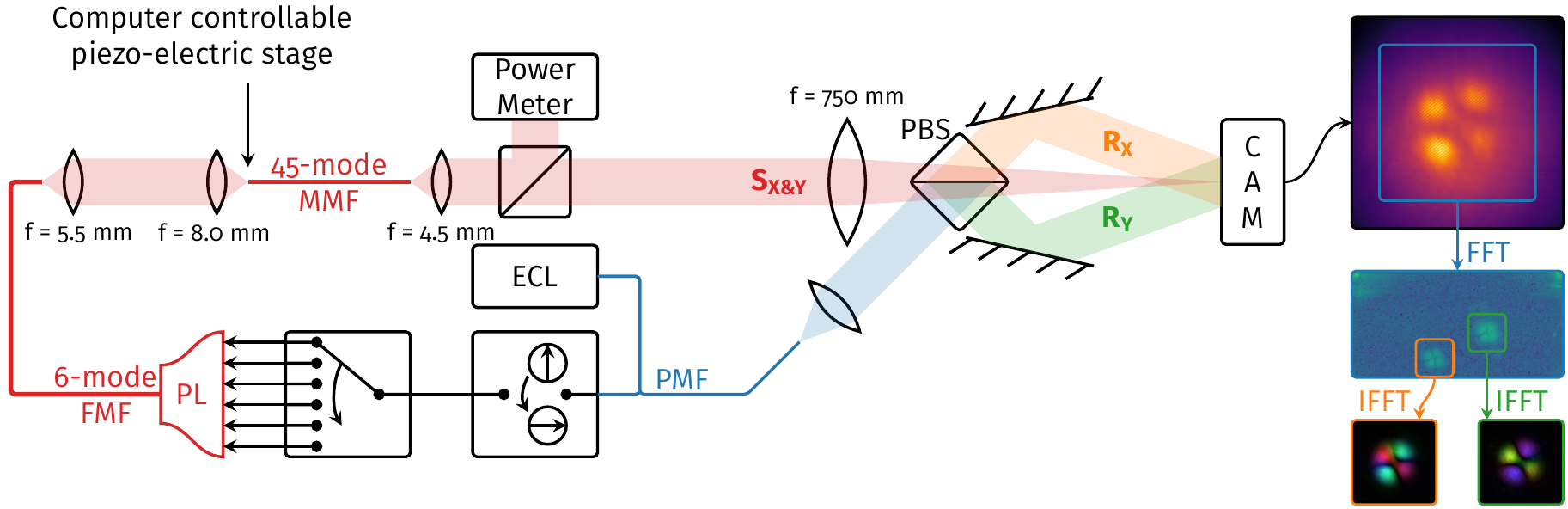}
    \caption{Experimental setup. PL: photonic lantern, S\textsubscript{X\&Y}: dual-polarization signal to be characterized, R\textsubscript{X} and R\textsubscript{Y}: reference beam for x- and y-polarization, FFT: fast Fourier transform, ECL: external cavity laser. Note that the signal S\textsubscript{X\&Y} passes over the PBS. }
    \label{fig:expsetup}
\end{figure*}
\cref{fig:expsetup} shows the experimental free-space optical setup. A \PL \cite{Velazquez-Benitez2018} multiplexes light from six single-mode fibers to a 6-mode \FMF \cite{sillar6LPFMF}. The light exiting the 6-mode \FMF is coupled into a short piece of 45-mode \MMF \cite{Sillard50um} in free space using collimator lenses mounted on computer controllable piezo-electric 3-axis stages. The light exiting the \MMF is collimated, split, and measured using a free-space power meter and off-axis \DH. The experimental \DH setup is comprised of a lens for the signal beam \textbf{S\textsubscript{X\&Y}}, a large collimator for the reference beams \textbf{R}, a \PBS, mirrors, and a camera. By interfering the signal field with two off-axis flat-phase reference beams, this setup is capable of measuring the signal field in both amplitude and phase for both polarizations. The digital field extraction process is visually explained in \cref{fig:expsetup} on the right. More details on the measurement technique, setup, and required \DSP can be found in \cite{thesisSjoerdDH,demoOFC}.

To optimize coupling, the x- and y-position of the 3-axis stage are swept. Coupling efficiency is measured using two methods. Firstly, light is inserted into one of the inputs of the \PL and the total optical power exiting the \MMF is measured using the free-space power meter. Secondly, the light exiting the \MMF is measured using off-axis \DH, providing a full description of the signal light which is used for subsequent modal analysis. The modal decomposition of the signal light is obtained through \textit{digital demultiplexing} into target mode fields, obtained for the employed \MMF using a scalar numerical mode solver. This process is repeated for each input port and polarization of the \PL to construct a complex-valued dual-polarization transfer matrix from \PL input port to \MMF output mode. Analysis of this transfer matrix reveals \MDL and mode-group \XT, which can be used to assess the quality of the transmission channel and free-space coupling therein.


\section{Results}

\begin{figure}[!t]
    \captionsetup[subfigure]{labelformat=empty}
    \centering
    \subfloat[\label{fig:poewr_lp01}]
    {
        \includegraphics[width=0.99\linewidth]{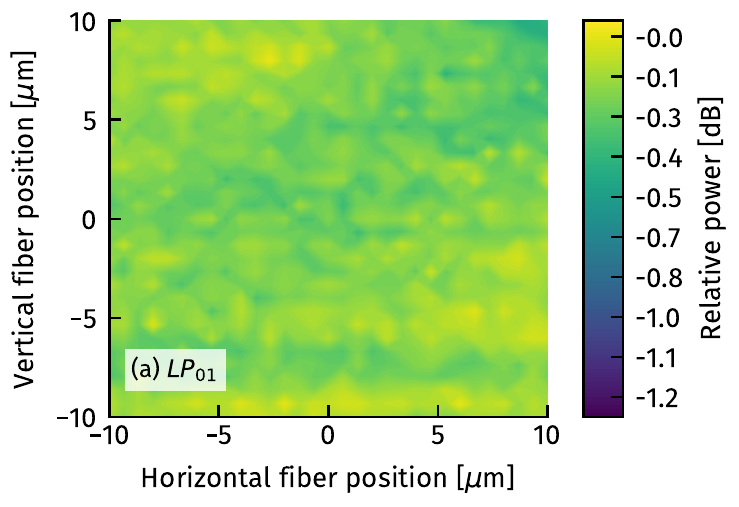}
    }\\ 
    \subfloat[\label{fig:power_lp02}]
    {
        \includegraphics[width=0.99\linewidth]{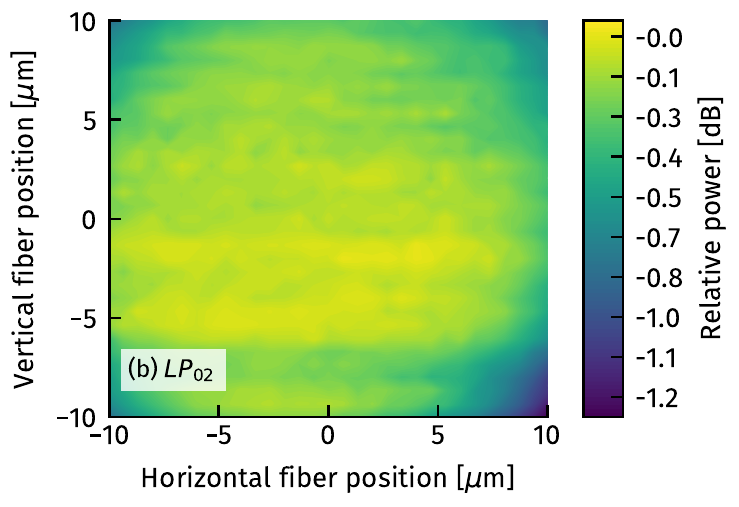}
    }
    \caption{Relative powers measured at output of the \MMF. Powers are normalized to the maximum measured powers. (a) shows the power when launching the \LP{01} and \LP{02} mode, respectively} 
    \label{fig:power}
\end{figure}

\cref{fig:power} shows the total coupled optical power measured using the free-space optical power meter when only the \LP{01} or \LP{02} port of the \PL is excited. As can be seen from this figure, no distinct optimum can be found using these power measurements. The coupled power stays constant within a range of \SI{0.5}{dB} when exciting the \LP{01} mode, while the fiber is scanned over a range of \SI{-10}{\micro\meter} to \SI{10}{\micro\meter} in both the horizontal and vertical direction. When exciting the \LP{02} mode using the \PL, the coupled power is about \SI{1.2}{dB} lower at large offsets, but constant at small offsets, similar to the \LP{01}. The slightly lower powers at large offsets are expected, as the \LP{02} mode couples with more difficulty into the \MMF compared to the \LP{01} mode, making it more suitable to assess quality of alignment. However, since no variation of the metric is observed over significant variation of the fiber position, it is not suitable for accurate alignment.

\begin{figure}
    \centering
    \includegraphics[width=0.99\linewidth]{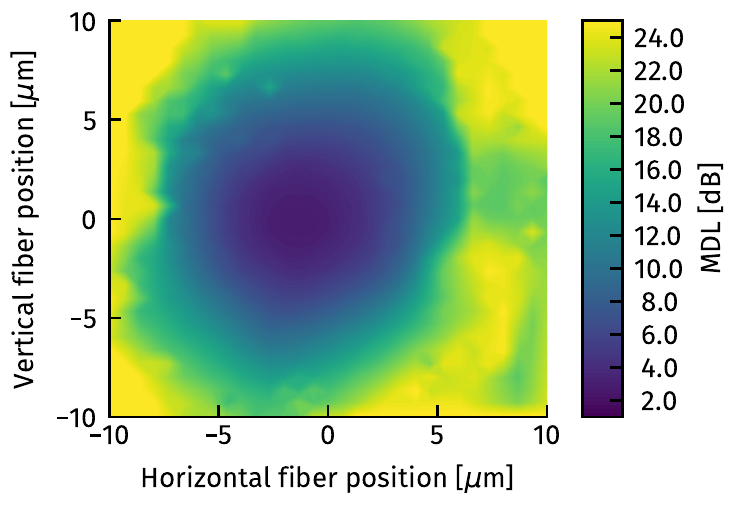}
    \caption{\MDL for different offsets of the 45-mode \MMF.}%
    \label{fig:mdl}
\end{figure}

\cref{fig:mdl} shows the \MDL measured using the \DH setup. Here \MDL is calculated through \SVD of the complex-valued dual-polarization transfer matrix $T$ from \PL input port to \MMF output mode:

\begin{align}
    \text{MDL\,[dB]} = 10 \cdot \log_{10} \left( \frac{\lambda_0}{\lambda_{2 N_k -1}} \right)^2
    \label{eq:mdl}
\end{align}
with $\lambda_0$ and $\lambda_{2 N_k -1}$ the largest and smallest singular value, respectively. The full transfer matrix $T$ can be converted into a mode-group intensity transfer matrix $\hat{T}$, from which \XT can be calculated using:

\begin{align}
    \text{XT\,[dB]} = 10 \cdot \log_{10} \left( \frac{\text{tr} (\hat{T})}{\Sigma \hat{T} - \text{tr} (\hat{T})} \right)
    \label{eq:xt}
\end{align}
with tr the trace operator. This definition of \XT describes the ratio between of power coupled to the intended mode-group and the other mode-groups. Thus, a higher value indicates more power on the diagonal of the matrix and less \XT to other mode-groups.

\begin{figure}
    \centering
    \includegraphics[width=0.99\linewidth]{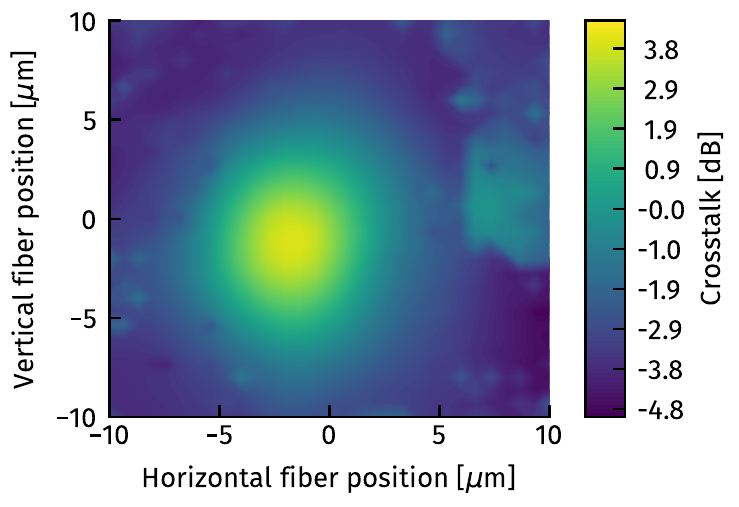}
    \caption{\XT for different offsets of the 45-mode \MMF.}%
    \label{fig:xt}
\end{figure}

A clear optimum fiber position is observed in \cref{fig:mdl} near zero offset in both horizontal and vertical direction. For fiber position offsets within the large optimum power area of \cref{fig:power_lp02}, \cref{fig:mdl} shows an \MDL penalty of up to \SI{20}{dB}, demonstrating that power measurements only do not guarantee adequate coupling. Furthermore, \cref{fig:xt} shows the measured \XT and a distinct optimal location for the fiber position is observed, which coincides with the optimum position obtained from \cref{fig:mdl}. Thus, both \MDL and \XT can be used as optimization metrics for alignment since they both directly measure the quality of coupling.

\section{Conclusions}
Optimization of free-space coupling between a \lFMF and a \lMMF is investigated. When only the total coupled optical power is maximized, the spatial distribution of the light is not taken into account and good coupling with low \lMDL and \lXT cannot be guaranteed. Off-axis \lDH provides a full description of optical fields and is demonstrated to provide relevant metrics for the investigated coupling scenario. The proposed method can be used for reliable automated alignment of \SDM components, devices, and subsystems enabling the effective coupling of amplifiers and multiplexers into transmission fibers.


\section{Acknowledgements}

Partial funding is from  the Dutch NWO Gravitation Program on Research Center for Integrated Nanophotonics (Grant Number 024.002.033), from the KPN-TU/e Smart Two program and from the Dutch NWO Visitor's program (Grant number 040.11.743).


\printbibliography

\vspace{-4mm}

\end{document}
